\documentclass[aps,prl,showpacs,superscriptaddress,amssymb,twocolumn]{revtex4-1}
\pdfoutput=1
\usepackage{eqnarray,amsmath,braket,physics}
\usepackage{natbib}
\usepackage{graphicx}
\usepackage{enumitem}
\usepackage{appendix}
\usepackage{color}
\usepackage[export]{adjustbox}
\usepackage{epstopdf} 
\DeclareGraphicsExtensions{.pdf,.eps,.png,.jpg,.mps}
\usepackage[colorlinks=true,citecolor=blue,urlcolor=blue,linkcolor=blue]{hyperref}
\usepackage[position=top,singlelinecheck=off,justification=raggedright,caption=false]{subfig}
\usepackage{bm,amsmath}
\usepackage{dcolumn}
\usepackage{multirow}
\usepackage{hyperref}
\usepackage{float}
\def\BEq{\begin{equation}}
\def\EEq{\end{equation}}
\def\BEqA{\begin{eqnarray}}
\def\EEqA{\end{eqnarray}}
\def\BW{\begin{widetext}}
\def\EW{\end{widetext}}
\usepackage{array}
\newcommand{\evall}{$E_{\mathrm{val}}$}

\newcommand{\estvall}{$E_{\mathrm{ST_{val}}}$}
\newcommand{\estorb}{$E_{\mathrm{ST_{orb}}}$}

\newcommand{\Si}{$\mathrm{S}$}
\newcommand{\Tvall}{$\mathrm{T_{val}}$}
\newcommand{\Torb}{$\mathrm{T_{orb}}$}

\begin{document}
\title{Charge-noise resilience of two-electron quantum dots in Si/SiGe heterostructures}

\author{H.\ Ekmel Ercan}
\affiliation{Department of Physics, University of Wisconsin-Madison, Madison, Wisconsin 53706, USA}
\author{Mark Friesen}
\affiliation{Department of Physics, University of Wisconsin-Madison, Madison, Wisconsin 53706, USA}
\author{S.\ N.\ Coppersmith}
\affiliation{Department of Physics, University of Wisconsin-Madison, Madison, Wisconsin 53706, USA}
\affiliation{School of Physics, The University of New South Wales,
Sydney, NSW 2052, Australia}

\date{\today}

\begin{abstract}
The valley degree of freedom presents challenges and opportunities for silicon spin qubits. 
An important consideration for singlet-triplet states is the presence of two distinct triplets, comprised of valley vs.\ orbital excitations.
Here we show that both of these triplets are present in the typical operating regime, but that only the valley-excited triplet offers intrinsic protection against charge noise. 
We further show that this protection arises naturally in dots with stronger confinement.
These results reveal an inherent advantage for silicon-based multi-electron qubits.
\end{abstract}    

\maketitle

When quantum dots contain more than one electron, new possibilities emerge for defining and controlling qubits.
Theoretical studies have shown that GaAs
dots with multiple electrons may be inherently protected from charge noise~\cite{Barnes,Mehl,BakkerMehl}, and recent experiments confirm some of these predictions~\cite{Kodera2009,Higginbotham2014,Chen2017}.
Recent progress in Si-based multi-electron qubits~\cite{hybridPRL,Kim2014,Harvey-Collard2017,Leon2020,Yang2020, Petit2020} brings renewed attention to such noise-reduction schemes.
However, the theoretical description of GaAs dots is not applicable to Si, due to the presence of both orbital and valley degrees of freedom for the electrons.  
While the orbital energies are determined by electrostatic confinement, similar to GaAs, the conduction-band valley splitting is determined by details of the quantum-well interface~\cite{Friesen2007}. 
It is technically challenging to describe such behavior because of the strong electron-electron (e-e) interactions that must be treated nonperturbatively, and because a minimal model of Si must include details of the Si band structure, as well as atomic-scale disorder at the quantum well interface, which gives rise to valley-orbit coupling (VOC)~\cite{Friesen2010}.

\begin{figure}
\includegraphics[width=8.6 cm]{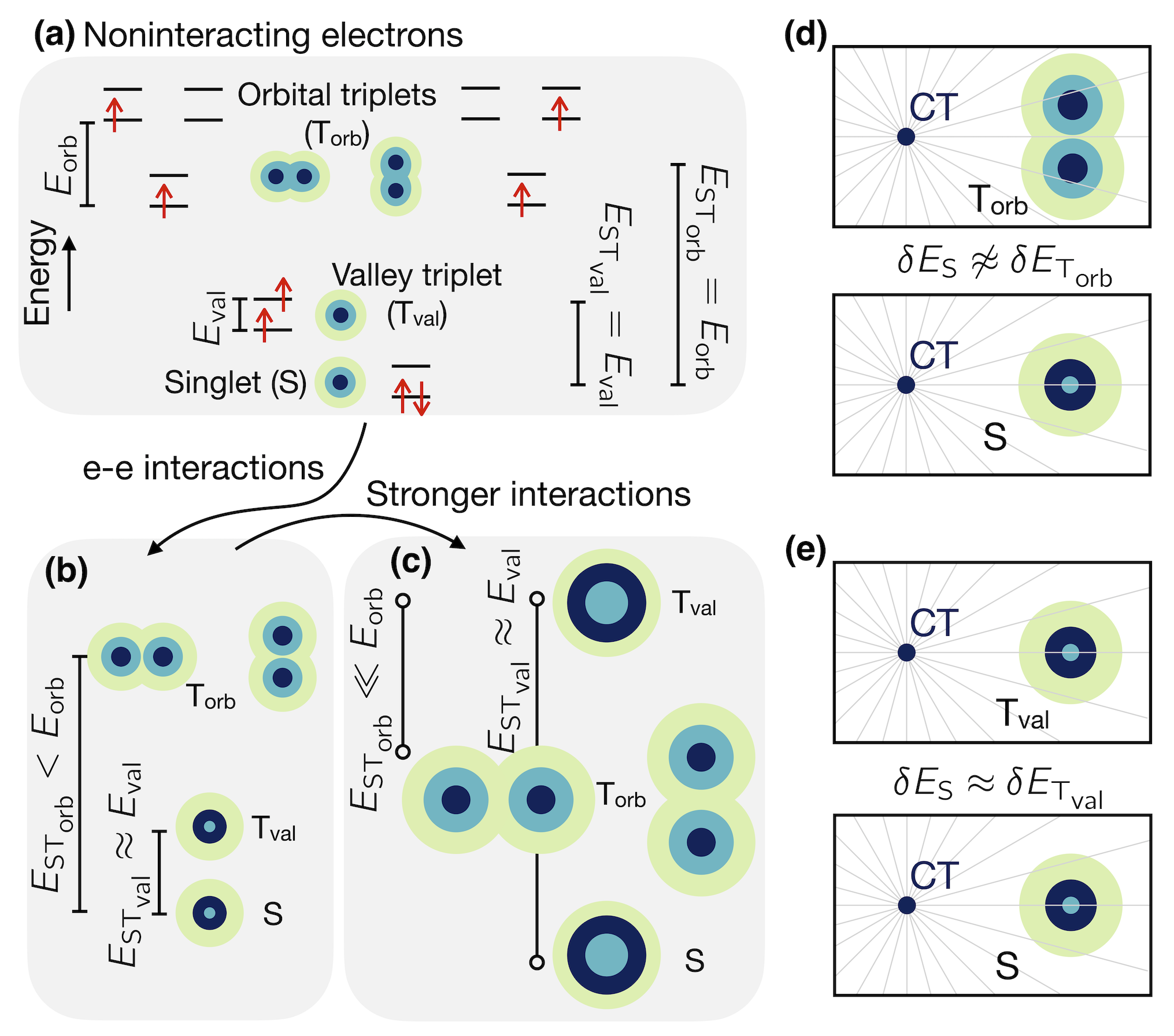}
\caption{Effects of e-e interactions and electrical noise on singlet-triplet states in a two-electron Si/SiGe quantum dot.
(a) Noninteracting electrons.
Single-electron energy levels (black lines) include valley and orbital excited states, with $E_\text{val}<E_\text{orb}$.
Two-electron states (S, T$_\text{val}$, T$_\text{orb}$) are formed from combinations of spin, valley, and orbital states (red arrows).
Charge distributions are shown schematically, with darker colors indicating higher densities.
The two T$_\text{orb}$ states have distinct $p_x$ or $p_y$ character.
(b) Including e-e interactions.
In the high-$E_\text{orb}$ regime, we find $E_{\text{ST}_\text{val}}<E_{\text{ST}_\text{orb}}$.
The resulting low-energy states (S and T$_\text{val}$) have very similar charge distributions.
(c) Stronger e-e interactions (low-$E_\text{orb}$ regime).
Here, $E_{\text{ST}_\text{orb}}<E_{\text{ST}_\text{val}}$, and the low-energy states (S and T$_\text{orb}$) have very different charge distributions.
(d),(e) The responses of \Si, \Torb, and \Tvall\ to a charge trap (CT) depend on their charge distributions. 
(d) Dissimilar distributions give large \estorb\ fluctuations. 
(e) Similar distributions give small \estvall\ fluctuations.}
\label{fig:fig1}
\end{figure}

Here we develop a complete theoretical toolbox for describing two-electron dots in Si.
We first apply these tools to study low-energy spin singlet and triplet states.
Solving the two-electron wavefunctions as a function of orbital confinement energy $E_\text{orb} \sim \hbar\omega$ reveals two fundamentally different triplet excitations, based on their valley or orbital character, as illustrated in Figs.~\ref{fig:fig1}(a)-\ref{fig:fig1}(c).
These excitations also have different coherence properties.
For small $\hbar\omega$, the low-energy states that define the qubit are the singlet (S) and orbital triplet (T$_\text{orb}$).
Since these states have dissimilar charge distributions, they couple differently to electrical fluctuations [e.g., a nearby charge trap (CT), as shown in Fig.~\ref{fig:fig1}(d)], resulting in dephasing. 
For stronger confinement (larger $\hbar\omega$), the low-energy states are S and the valley triplet (T$_\text{val}$). 
In this case, the charge distributions are very similar, and they respond similarly to electrical fluctuations [Fig.~\ref{fig:fig1}(e)], yielding qubits that are resilient to dephasing.

\begin{figure}
\includegraphics[width=8.6 cm]{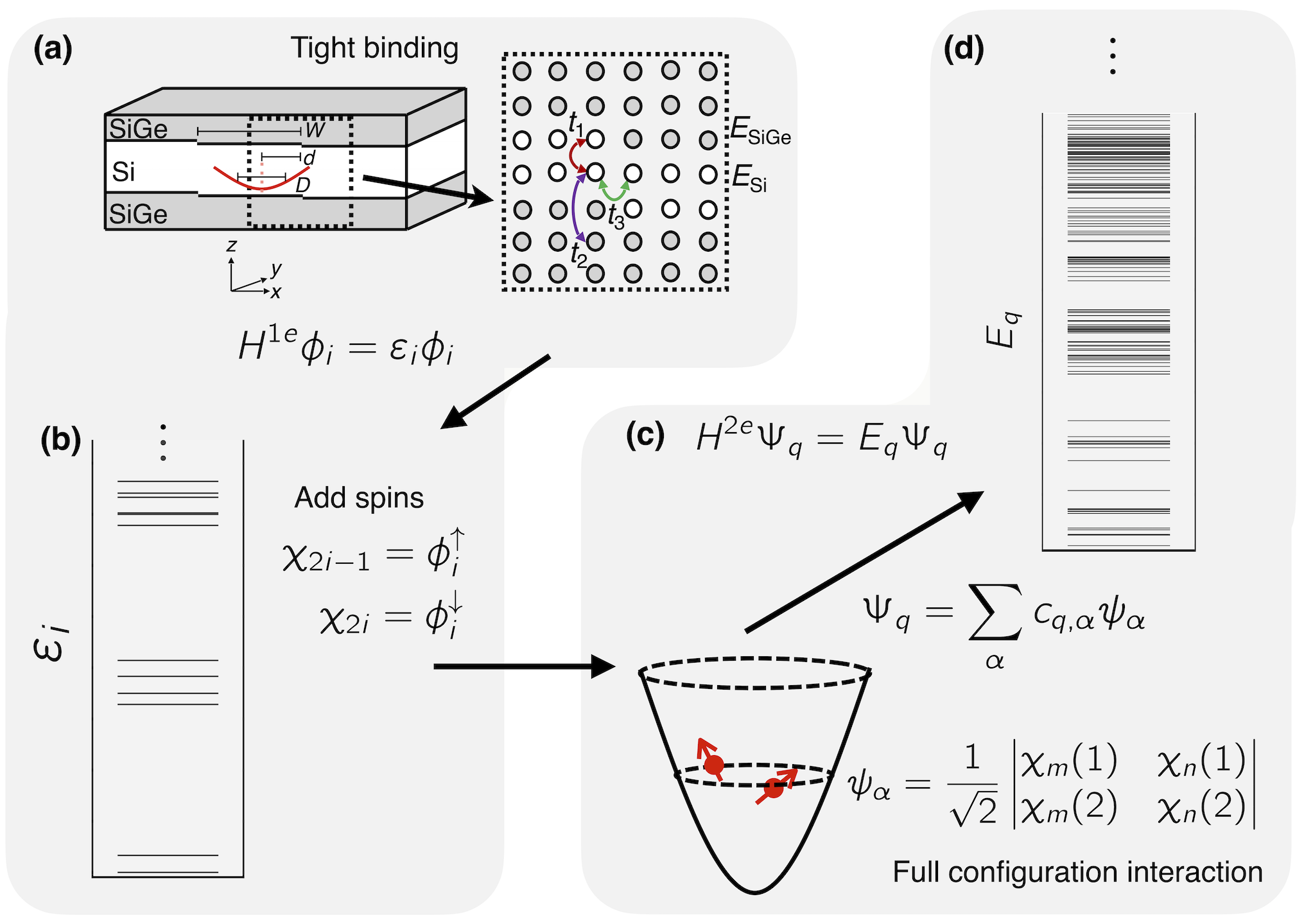}
\caption{Overview of theoretical methods.  
(a) Schematic of the 2D TB method used to compute single-electron wavefunctions, while accounting for atomic-scale disorder at the quantum well interface.
Si sites are shown as white and SiGe sites are shown as gray.
Interface steps have width $W$, the harmonic dot confinement potential (red) has diameter $D$, and we take the dot to be centered halfway between two steps, $d=W/2$.
Hopping parameters $t_1$, $t_2$, and $t_3$ and on-site parameters are discussed in the main text.
(b) Typical results for single-electron energies $\varepsilon_i$. 
(c) FCI step: Slater determinants $\psi_{\alpha}$ are computed for spin-orbitals $\chi_i$, obtained from TB valley-orbitals $\phi_i$, combined with spin coordinates. 
The two-electron Hamiltonian $H^{2e}$ is diagonalized in this Slater basis, with enough spin-orbit basis states (84) to ensure convergence. 
(d) Typical results for two-electron energies $E_q$. 
}
\label{fig:fig2}
\end{figure}

\emph{Theoretical methods.}
We compute two-electron wavefunctions in two steps.
First, we use a tight-binding (TB) approach to obtain single-electron wavefunctions~\cite{boykin_2004,boykin_2004_2}.
This method accounts for the essential features of the Si bandstructure and allows an atomistic description of disorder at the quantum well interface.
Second, we incorporate these single-electron wavefunctions into a full configuration interaction (FCI) \cite{szabo&ostlund} scheme for computing two-electron wavefunctions, nonperturbatively, while accounting for strongly interacting electrons.
The full method is summarized in Fig.~\ref{fig:fig2}; additional details are given in Ref.~\cite{Ekmel}.

In step one, the single-electron Hamiltonian for the 3D heterostructure is assumed to be separable in terms of the $(x,z)$ vs.\ $y$ variables, where $\hat x$, $\hat y$, and $\hat z$ are the crystallographic axes.
We consider atomistic disorder only in the $x$-$z$ plane.
This simplification allows us to treat the $(x,z)$ variables using TB methods, while solving the separable $y$ wavefunctions using continuum effective-mass theory, which allows us to achieve full convergence on a more practical time scale. 
The single-electron Hamiltonian can then be written as 
\begin{equation}
H^{\mathrm{1e}}=H_{\mathrm{K}}+H_{\mathrm{E}}+H_{\mathrm{QW}}.
\end{equation}
Here, the kinetic energy is given by
\begin{multline}
H_{\mathrm{K}}=\frac{-\hbar^2}{2m_t} \frac{\partial^2}{\partial y^2}  +\sum_{i_x,i_z} (t_1 \ket{i_x,i_z+1} \bra{i_x,i_z} \\ + t_2 \ket{i_x,i_z+2} \bra{i_x,i_z} + t_3 \ket{i_x+1,i_z}\bra{i_x,i_z} + \text{h.c.}) ,
\end{multline}
where the integer indices $i_x$ and $i_z$ refer to TB sites along the $\hat x$ and $\hat z$ axes, respectively.
We have suppressed the spin index here, because our Hamiltonian is independent of spin, and the effects of Pauli exclusion become important only at the FCI stage of the calculation. 
The hopping parameters $t_1=0.68$ eV and $t_2=0.61$~eV are chosen to reproduce the key features of the two-fold degenerate structure at the bottom of the Si conduction band: (1) valleys centered at $\mathbf{k}=\pm k_0 \hat z$ in reciprocal space, where $k_0=\pm 0.82 (2\pi/a)$, $a=5.43$~$\text{\normalfont\AA}$ is the cubic lattice constant, and $\Delta z=a/4$ is the grid spacing,
and (2) longitudinal effective mass, $m_l=0.916 \ m_0$.
The hopping parameter $t_3=-0.026$~eV gives the correct transversal effective mass, $m_t=0.191\, m_0$, for the grid spacing $\Delta x=2.79$~nm~\cite{JC2018}. 
We note that $\Delta x$ can be much larger than $\Delta z$, because there are no fast valley oscillations along $\hat x$.
The vertical (quantum well) confinement potential is given by
\begin{multline}
H_{\mathrm{QW}}=\sum_{i_x,i_z} \Big[E_0+V_{\mathrm{QW}} \Theta_{i_x,i_z} \\ - e(i_x F^\text{e}_x \Delta x+i_z F^\text{e}_z  \Delta z)\Big] \ket{i_x,i_z} \bra{i_x,i_z},
\end{multline}
where $\Theta_{i_x,i_z}$ is a step function that takes the value $1$ on a SiGe site and $0$ on a Si site, $V_{\mathrm{QW}}=150$~meV is the band offset between Si and SiGe, $e=|e|$ is the elementary charge, and $\bm{F}^{\mathrm{e}}=(F^{\mathrm{e}}_x,0,F^{\mathrm{e}}_z)$ is the electric field perpendicular to the interface due to the gate electrodes. 
All calculations assume 10~nm quantum wells. Interface disorder is implemented via the choice of $\Theta_{i_x,i_z}$. 
In this work we consider an interface tilted slightly away from $\hat z$, with uniformly-distributed single-atom steps of height $a/4$ and width $W$, as depicted in Fig.~\ref{fig:fig2}(a). 
The field $\bm{F}^{\mathrm{e}}$ is taken to be perpendicular to the tilted interface.
The lateral confinement potential is of electrostatic origin, and is taken to be parabolic, with the form
\begin{multline}
H_{\mathrm{E}}=\frac{1}{2}m_t \omega_x^2 \sum_{i_x,i_z} (i_x \Delta x)^2 \ket{i_x,i_z} \bra{i_x,i_z} + \frac{1}{2}m_t \omega_y^2 y^2.
\label{eq:HE}
\end{multline}
We solve $H^{\mathrm{1e}} \phi_i=\varepsilon_i \phi_i$ to obtain the single-electron basis states $\phi_i$ used in the FCI calculation. 
A typical energy level structure is shown in Fig.~\ref{fig:fig2}(b).

In step two, we solve the two-electron Hamiltonian, which includes the Coulomb interaction term,
\begin{equation}
H^{\mathrm{2e}}=H^{\mathrm{1e}}(\bm{r}_1)+H^{\mathrm{1e}}(\bm{r}_2)+\frac{e^2}{4 \pi \epsilon_0 \epsilon_r } \frac{1}{\abs{\bm{r}_1-\bm{r}_2}}~,
\label{eq:H2e}
\end{equation} 
where $\epsilon_0$ is the permittivity of free space, and
$\epsilon_r=11.4$ is the dielectric constant of low-temperature Si~\cite{Faulkner}.
The Coulomb matrix elements are computed using a combination of numerical and analytical methods, taking advantage of the analytical wavefunction solutions along $\hat y$.
We then solve for the eigenvalues and eigenstates of $H^{\mathrm{2e}} \Psi_q=E_q \Psi_q$ using FCI methods: $H^{\mathrm{2e}}$ is diagonalized in a basis of Slater determinants generated from spin orbitals $\chi=\phi\,\otimes\uparrow$ or $\chi=\phi\,\otimes\downarrow$, as shown in Figs.~\ref{fig:fig2}(c) and \ref{fig:fig2}(d). 
A finite set of determinants is used and the convergence of the FCI method is checked, as in Ref.~\cite{Ekmel}

\begin{figure*}
\includegraphics[width=17.2 cm]{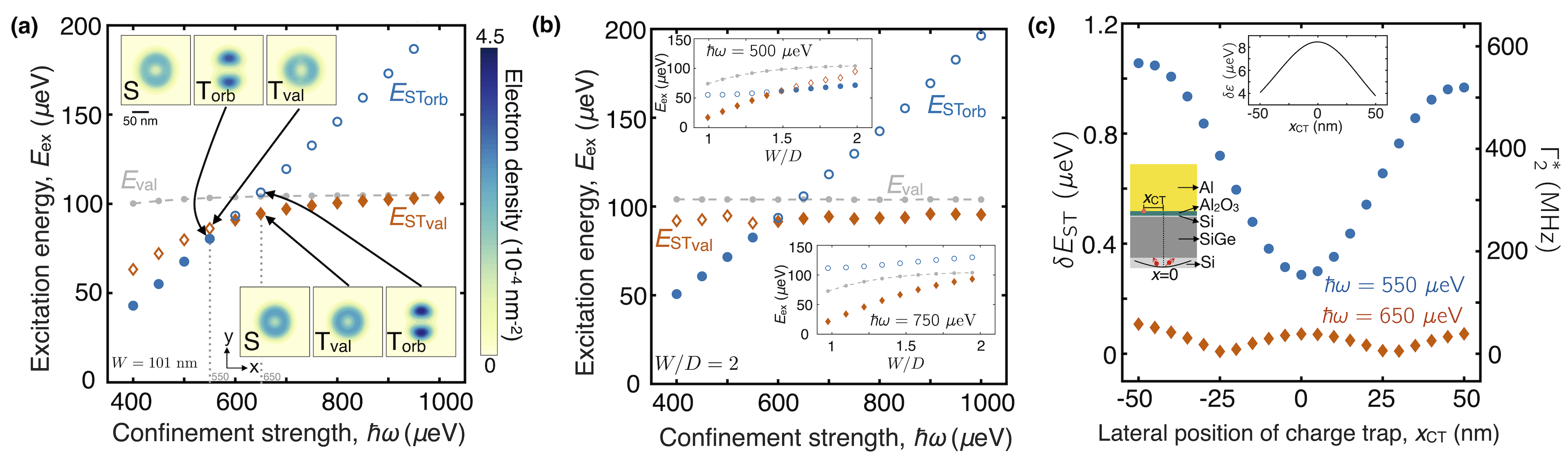}
\caption{Effects of e-e interactions, interface steps, and a charge trap on the excitation energies of a two-electron dot.
Solid symbols refer to the lowest ST excitation, which defines the qubit, and open symbols  refer to the higher ST excitation.
(a) ST splittings for fixed terrace width, $W=100$~nm, with the dot center equidistant between two steps.
 [See Fig.~\ref{fig:fig2}(a) for device geometry.]
 The single-electron valley splitting \evall\ is also shown.  
 $E_{\text{ST}_\text{orb}}$ is strongly suppressed below $E_\text{orb}=600$~$\mu$eV (not shown) over its entire range, due to strong e-e interactions.
 A crossover is observed between regions dominated by ST$_\text{orb}$ vs.\ ST$_\text{val}$.
 Here, $E_{\text{ST}_\text{orb}}$ is typically too small to form practical qubits in the small-$\hbar\omega$ regime.  
Insets: charge densities of S, \Torb, and \Tvall\ states, for two different confinement strengths. 
(b) ST splittings for fixed $W/D=2$, where the dot diameter $D$ depends on $\hbar\omega$ (so $W$ also depends on $\hbar\omega$).
Here, $E_{\text{ST}_\text{val}}\approx E_\text{val}$ is approximately constant, indicating that VOC is mainly determined by the overlap of the wavefunctions with interface steps.
Insets: the same quantities are plotted as a function of $W/D$ for fixed $\hbar\omega$, showing that $E_{\text{ST}_\text{val}}$ is more strongly affected by the steps than $E_{\text{ST}_\text{orb}}$.
In (a) and (b), the triplet crossover occurs in an experimentally relevant regime.
(c) Shift in $E_{\mathrm{ST}}$, and corresponding qubit dephasing rate $\Gamma^*_2$, as a function of lateral trap position, $x_\text{CT}$, defined in the lower inset. 
Results are only shown for the low-lying excitations, just below or above the triplet crossover in (a).
$\Gamma_2^*$ is significantly lower for qubits defined by \Tvall\, since S and T$_\text{val}$ have nearly identical charge distributions.
Upper inset: 
shift of the detuning $\varepsilon$ of a double dot, for dots separated by 200~nm, due to the occupation of a charge trap at lateral position $x_\text{CT}$. 
}
\label{fig:fig3}
\end{figure*}

{\it Results.} 
To better understand the nontrivial behavior arising from e-e interactions and VOC, it is instructive to consider these effects one at a time, as summarized in Fig.~\ref{fig:fig1}.
In the absence of interactions, the ground-state singlet (S) is formed from two electrons, each in the lowest orbital level.
There are two different types of single-electron excitations above the ground state: valleys and orbitals, with corresponding single-electron excitation energies, $E_\text{val}$ and $E_\text{orb}$.
Two-electron excitations take on the character of these single-electron excitations, yielding distinct valley (T$_\text{val}$) and orbital (T$_\text{orb}$) triplets, with excitation energies $E_{\text{ST}_\text{val}}=E_\text{val}$ and $E_{\text{ST}_\text{orb}}=E_\text{orb}$.
Here we assume $E_\text{val}<E_\text{orb}$, as consistent with many qubit experiments~\cite{valleyMeasure1,valleyMeasure2,valleyMeasure3,chen,Zajac,JP,corrigan}. 
The charge distribution of T$_\text{val}$ is identical to S; however T$_\text{orb}$ is quite different, due to its $p$-orbital contribution.
Now, introducing e-e interactions [Fig.~\ref{fig:fig1}(b)], we find that many additional Slater determinants contribute to the two-electron wavefunctions.
To a very good approximation, the triplets retain their valley or orbital character; however the e-e interactions strongly suppress $E_{\text{ST}_\text{orb}}$ below $E_\text{orb}$, while having almost no effect on $E_{\text{ST}_\text{val}}\approx E_\text{val}$~\cite{Ekmel}.
The charge distributions for S and T$_\text{val}$ take the form of Wigner-molecule doughnuts, with dimples at their centers~\cite{Bryant,Yanno,Egger,Filinov,Grabert}.
For stronger interactions [i.e., smaller $\hbar\omega$, Fig.~\ref{fig:fig1}(c)], there is a crossover from T$_\text{val}$ to T$_\text{orb}$-dominated excitations.
In this regime, the qubit states (S and T$_\text{orb}$) have very different charge distributions.

We finally consider a realistic dot model including e-e interactions and VOC.
Typical results for excitation energies are shown in Fig.~\ref{fig:fig3}(a).
To begin, we consider wide steps, $W=101$~nm, to clearly demonstrate the types of behavior observed as a function of $\hbar\omega$.
We also position the dot as far as possible from a step, with $d=W/2$, as depicted in Fig.~\ref{fig:fig2}(a).
For small $\hbar\omega$ (weak confinement), T$_\text{orb}$ is the dominant excitation, with an excitation energy, $E_{\text{ST}_\text{orb}}\lesssim k_BT$, that is typically too small to enable high fidelity qubit initialization or readout.
For larger $\hbar\omega>600$~$\mu$eV, there is a crossover to T$_\text{val}$-dominated behavior.
If the valley splitting is also $\gtrsim 100$~$\mu$eV, the energy $E_{\text{ST}_\text{val}}$ will certainly be large enough for practical applications.
(For quantum-dot hybrid qubits, slightly smaller $E_{\text{ST}_\text{val}}$ are preferred~\cite{hybridPRL,Kim2014}.)
For this calculation, we used $\abs{\bm{F}^{\mathrm{e}}}$=0.6~MV/m, which gives $E_\text{val}\sim 105$~$\mu$eV.
For $\hbar\omega$ values below the triplet crossover, \estvall\ drops quickly, as the dot (with diameter $D= 2 \sqrt{ \hbar / m_t  \omega}$) begins to strongly overlap with different interface steps, suppressing the valley splitting~\cite{Friesen2007}, and causing VOC~\cite{Friesen2010}.
The value of $W=2D$ used in Fig.~\ref{fig:fig3}(a) was chosen such that the ST splitting is not significantly affected by VOC.
We can also explore other regimes by computing the excitation energies as a function of $\hbar\omega$ while holding $W=2D$ fixed, as shown in Fig.~\ref{fig:fig3}(b), to ensure that the dot does not interact excessively with the steps.
(This requires simultaneously changing $W$ as $\hbar\omega$ is varied.)
Here, we again observe a crossover from T$_\text{orb}$ to T$_\text{val}$-dominated behavior.
However, because the wavefunction remains far from the nearest step, \estvall\ is not suppressed for low $\hbar\omega$.
In contrast, the insets of Fig.~\ref{fig:fig3}(b) show that smaller $W/D$ ratios cause significant reductions in ST splittings, making it more difficult to achieve acceptable values for qubit applications. 
In the Supplementary Materials we further discuss interface profiles that give small ST splittings.

The crossover from T$_\text{orb}$ to T$_\text{val}$-dominated behavior has a strong effect on qubit coherence, because the T$_\text{orb}$ and T$_\text{val}$ charge distributions couple differently to charge fluctuations. 
For  the quantum-dot hybrid qubit~\cite{hybridPRL,Kim2014}, for example, $E_\text{ST}$ determines the qubit energy, and fluctuations of $E_\text{ST}$ lead directly to dephasing~\cite{Brandur}.
The insets of Fig.~\ref{fig:fig3}(a) show typical in-plane electron densities for $\hbar \omega=550$ and 650~$\mu$eV, which bracket the crossover between low-energy triplet states. 
As noted above, the charge distribution of S is very similar to that of T$_\text{val}$ but not T$_\text{orb}$.
These distinctions are still valid when the VOC, which mixes the valley and orbital character of the wavefunctions, is weak but nonzero.
Consequently, charge noise affects S and T$_\text{val}$ similarly, yielding weak fluctuations of \estvall, but much larger fluctuations of \estorb. 
The high-$\hbar\omega$ regime is therefore expected to yield qubits with much better coherence properties.

To quantify these claims, we consider the effect of a charge trap, as depicted in Figs.~\ref{fig:fig1}(d) and \ref{fig:fig1}(e).
First-order perturbation theory is used to estimate the shifts in $E_\text{ST}$ due to the electrostatic potential $V_\text{CT}$ of the trap,
\begin{equation}
\delta E_{\mathrm{ST_{val(orb)}}} \approx e \left| \bra{\mathrm{T_{val (orb)}}} V_\text{CT} \ket{\mathrm{T_{val(orb)}}}-\bra{\mathrm{S}} V_\text{CT} \ket{\mathrm{S}} \right| .
\label{Eq:dSTval}
\end{equation}
We note that interfacial disorder breaks the circular symmetry, allowing us to use nondegenerate perturbation theory for the circular confinement potentials used in this work.
We evaluate Eq.~(\ref{Eq:dSTval}) for the geometry shown in the inset of Fig.~\ref{fig:fig3}(c), assuming a 10~nm Si quantum well, a SiGe barrier of width 40~nm, a 1~nm Si cap, a 5~nm layer of $\mathrm{Al_2O_3}$, and a metal top gate.
For simplicity, we consider the gate to be an infinite plane giving rise to a uniform electric field, $\bm{F}^{\mathrm{e}}$, but not the dot confinement potential.
The dot confinement is simply given by Eq.~(\ref{eq:HE}), and we assume the image potentials for the dot electrons are subsumed into this potential.
We take the charge trap to be located $\sim$50 nm above the dot, inside the oxide layer, as suggested by recent experiments~\cite{Connors2019}. 
Due to its proximity to the top gate, the trap is strongly screened.
Following Ref.~\cite{takashima1978}, and using the dielectric constants of the different layers, we obtain the leading terms in the potential,
$V_\text{CT}\approx\frac{1.13e}{4 \pi \epsilon_0 \epsilon_r }\left(\abs{\bm{R}-\bm{r}}^{-1}-\abs{\bm{R}_\text{im}-\bm{r}}^{-1}\right)$, where $\bm{R}$ is the position of the trap, $\bm{R}_\text{im}$ is the position of its mirror image inside the metal, and $\bm{r}$ is the position of the dot. 
To estimate the distance between the trap and the top gate, we consider a double-dot geometry with an interdot separation of 200~nm.
The top inset of Fig.~\ref{fig:fig3}(c) shows the shift $
\delta \varepsilon$ in the double-dot detuning parameter $\varepsilon$, caused by a charge trap separated from the dot by a lateral distance $x_\text{CT}$.
For a trap located 0.1~nm below the gate, the resulting shifts fall into the range 4-9~$\mu$eV, as consistent with experimental measurements of detuning fluctuations $\sigma_{\varepsilon}$~\cite{Wu2014,Brandur,Mi2018,Watson2018}.

Our perturbative results for the dominant ST splittings are shown in Fig.~\ref{fig:fig3}(c), as obtained on either side of the triplet crossover, at locations $\hbar \omega=550$~$\mu$eV (ST$_\text{orb}$) or $\hbar \omega=650$~$\mu$eV (ST$_\text{val}$).
Here, $x_\text{CT}=0$ corresponds to a charge trap located directly above the dot.
As expected, the energy fluctuations are strongly suppressed for ST$_\text{val}$.
We can also estimate dephasing rates for a double-dot qubit from the relation $\Gamma^*_2=\frac{\sigma_{\varepsilon}}{\sqrt{2} \hbar}\abs{\partial E_Q / \partial \varepsilon} $~\cite{dial,petersson2010,JC2018}.
Assuming the charge trap has equal switching rates between its empty and occupied states, we can approximate the standard deviation of $E_{\mathrm{ST_{val(orb)}}}$ fluctuations as $(1/2)\delta E_{\mathrm{ST_{val(orb)}}}$, so that $\sigma_{\varepsilon} \abs{\partial E_Q / \partial \varepsilon} \approx (1/2) \delta E_{\mathrm{ST_{val(orb)}}}$.
Dephasing estimates obtained in this way are also reported in the main panel of Fig.~\ref{fig:fig3}(c).
For the settings considered here, we see that the dephasing rates for ST$_\text{val}$ vs.\ ST$_\text{orb}$ can differ by a very large factor ($\sim 10$), depending on the lateral position of the trap. 
The lower curve, associated with \estvall, appears to be more consistent with recent experimental measurements of $\Gamma^*_2=6$-210~MHz in a Si hybrid qubit~\cite{Brandur}.
These results may also help to explain the much higher dephasing rates observed in a GaAs hybrid qubit~\cite{Cao:2016p086801}, $\Gamma^*_2=0.12$-1.4~GHz, which has no T$_\text{val}$ state.
(Note that the current results are obtained using Si materials parameters.)

{\it Summary.} 
Using a combination of tight-binding and full-configuration-interaction calculations,
we have shown that an important crossover occurs in the low-lying triplet state of two-electron dots in Si/SiGe: for weak confinement,
the orbital triplet is the dominant excitation, while for strong confinement the valley triplet is dominant.
We find that strong e-e interactions and valley-orbit coupling (induced by atomic steps at the quantum-well interface) both play key roles in this behavior, in the physically relevant operating regime.
We further show that the charge distribution of the valley triplet is similar to that of the singlet, but differs from the orbital triplet.
Consequently, qubits based on valley-triplet excitations are much more resilient to charge noise.
These results are crucial for successful implementations of multi-electron qubits in Si/SiGe dots.

\begin{acknowledgments}
We thank Mark Eriksson, J.\ P.\ Dodson, Joelle Corrigan, Tom McJunkin, Leah Tom, Merritt Losert, Jos\'{e} Carlos Abadillo-Uriel, Mark Gyure, Samuel Quinn, Andrew Pan, Joseph Kerckhoff, Elliot Connors, and John Nichol for helpful discussions.
This work was supported in part by ARO through Award No.\ W911NF-17-1-0274 and the Vannevar Bush Faculty Fellowship program sponsored by the Basic Research Office of the Assistant Secretary of Defense for Research and Engineering and funded by the Office of Naval Research through Grant No.\ N00014-15-1-0029. The views and conclusions contained in this document are those of the authors and should not be interpreted as representing the official policies, either expressed or implied, of the U.S. Government. The U.S. Government is authorized to reproduce and distribute reprints for Government purposes notwithstanding any copyright notation herein. This research was performed using the compute resources and assistance of the UW-Madison Center For High Throughput Computing (CHTC) in the Department of Computer Sciences. The CHTC is supported by UW-Madison, the Advanced Computing Initiative, the Wisconsin Alumni Research Foundation, the Wisconsin Institutes for Discovery, and the National Science Foundation, and is an active member of the Open Science Grid, which is supported by the National Science Foundation and the U.S. Department of Energy's Office of Science.
\end{acknowledgments}

\renewcommand\thefigure{\thesection.\arabic{figure}}    
\renewcommand{\thesection}{S\arabic{section}}  
\renewcommand{\thesubsection}{S\arabic{subsection}}  
\renewcommand{\thetable}{S\arabic{table}}  
\renewcommand{\thefigure}{S\arabic{figure}}
\renewcommand{\figurename}{FIG.}

\section{Supplementary material}
\setcounter{figure}{0}   
\setcounter{subsection}{1}   
\subsection{S\arabic{subsection}. Singlet-triplet splitting in the presence of strong valley-orbit coupling}
The main text describes the results of calculations performed in the regime where the valley-orbit coupling (VOC) induced by atomic steps at the quantum well interface is small enough that the singlet-triplet splitting is of order $100$~$\mu$eV, which is large enough for quantum computing.
Figure~\ref{fig:fig3} shows results obtained when the dot is maximally separated from the nearest step, corresponding to $d=W/2$ in Fig.~\ref{fig:fig2}(a), where $d$ is the distance from the dot center to the nearest step and $W$ is the distance between steps, or terrace width.
However, VOC is enhanced when the dot is located closer to a step.
Here we examine such cases of enhanced VOC, by characterizing the single-electron valley splitting and the two-electron singlet-triplet energy splittings for different dot-step separations.

\begin{figure}[t]
\centering
\includegraphics[width=8.6 cm]{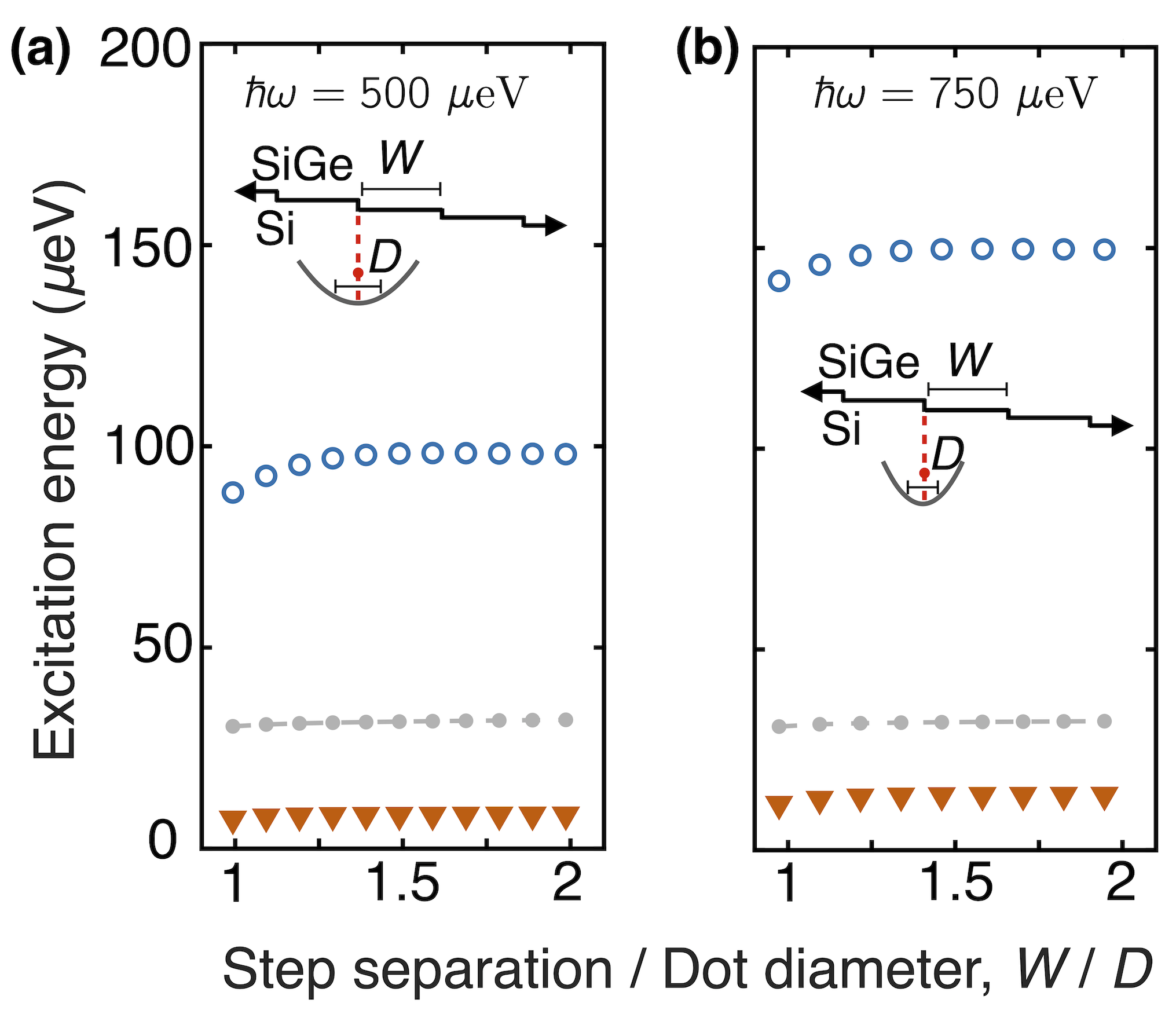}
\caption{
ST and valley splittings for a dot centered above a step (see inset), with $\hbar\omega$ (and therefore the dot diameter $D$) held constant.
(a) $\hbar\omega=500$~$\mu$eV.
(b) $\hbar\omega=750$~$\mu$eV.
All other parameters are the same as the insets of Fig.~\ref{fig:fig3}(b), where the dot is centered halfway between steps.
From top to bottom: $E_{\text{ST}_\text{orb}}$ (blue), $E_\text{val}$ (gray), and $E_{\text{ST}_\text{val}}$ (red).
When no steps are present, these same parameters yield $E_\text{val}\approx 100$~$\mu$eV.
We see that the valley splitting $E_\text{val}$ is suppressed by the steps, and that the valley singlet-triplet splitting $E_{\text{ST}_\text{val}}$ is additionally suppressed to be much less than $E_\text{val}$ by e-e interactions.
Note that $E_\text{val}$ and $E_{\text{ST}_\text{val}}$ are both nearly independent of $W/D$, indicating that the main effect on valley splitting and VOC is from the step under the dot.
}
\label{fig:figS1}
\end{figure}

Figures~\ref{fig:figS1}(a) and ~\ref{fig:figS1}(b) show results for the valley splitting and the ST splittings, for both types of triplet excitations, as a function of $W/D$.
Here, the dots are centered directly above a step, which maximizes the VOC.
The results can be directly compared to the insets in Fig.~\ref{fig:fig3}(b), which are obtained for the same device parameters, but with the dot located halfway between two steps, which minimizes VOC.
The differences between the two figures are striking.
First, in Fig.~\ref{fig:figS1}, we see that the single-electron valley splitting (dashed line) is appreciably reduced, as is well known for dots near a step~\cite{Friesen2007}.
Additionally, the strong enhancement of VOC allows e-e interactions to strongly suppress $E_{\text{ST}_\text{val}}$, while in some cases enhancing $E_{\text{ST}_\text{orb}}$~\cite{Ekmel}.
This leads to a qualitative change in behavior, for which there is no longer a crossover between the two triplet states as a function of $\hbar \omega$.
For any physically realistic $\hbar\omega$ value, when a dot is centered directly above a step, T$_\text{val}$ is the low-energy excited state, and its excitation energy is too small to allow high-fidelity single-electron qubits.

\begin{figure}[t]
\centering
\includegraphics[width=2.9in]{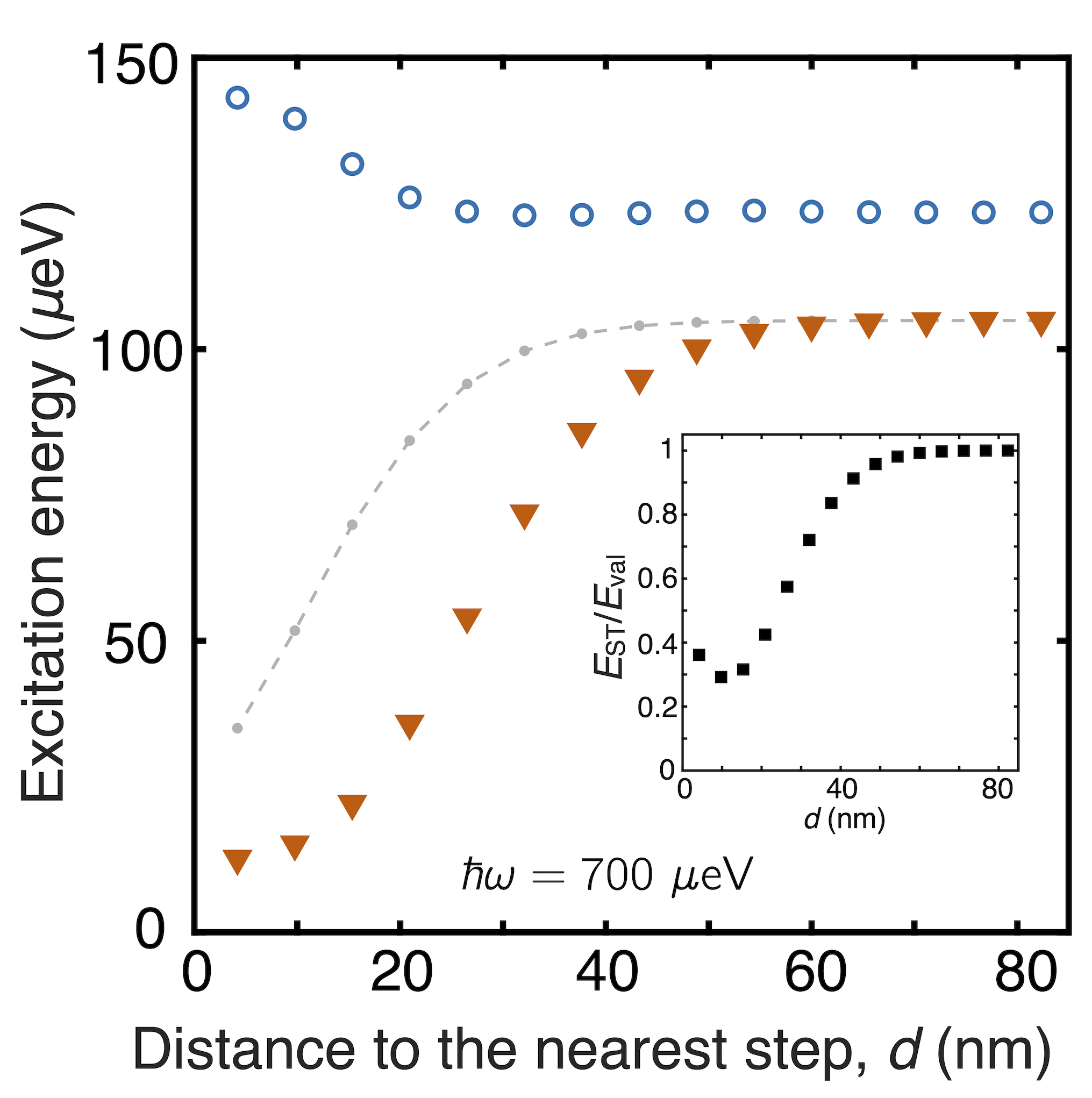}
\caption{
Dependence of the excitation energies on dot-step separation, $d$, when $W/D=3.5$ and $\hbar\omega = 700$~$\mu$eV are held fixed. 
Top to bottom: $E_{\text{ST}_\text{orb}}$ (blue), $E_\text{val}$ (gray), and $E_{\text{ST}_\text{val}}$ (red).
We observe a crossover from strong suppression of $E_\text{val}$ and $E_{\text{ST}_\text{val}}$, with $E_{\text{ST}_\text{val}}\ll E_\text{val}$, to almost no suppression, with $E_\text{val}\approx E_{\text{ST}_\text{val}}$.
Inset: interdependence of VOC and e-e interactions induces nonmonotonic variations of $E_\text{ST}/E_\text{val}$.
}
\label{fig:figS2}
\end{figure}

In Fig.~\ref{fig:figS2}, we plot the valley splitting and ST splittings for fixed $\hbar \omega =700 \ \mu$eV and step separation $W=3.5D$,
as a function of the dot-step separation $d$. 
When $d \ll D$, $E_\text{val}$ and $E_{\text{ST}_\text{val}}$ are strongly suppressed, as in Fig.~\ref{fig:figS1}.
However energy splittings increase as the dot-step separation $d$ increases.
For this particular case, we see that lateral dot motion on the order of 30-50~nm allows us to achieve robust two-electron qubits.
For single-electron qubits, the transition to robust qubit energies occurs at lower dot-step separations, of order 20~nm, due to the absence of e-e interactions.
Finally, we note that the dependence of the two-electron excitation energies on valley splitting and e-e interactions can be complicated, as demonstrated by the nonmonotonic dependence of $E_\text{ST} / E_\text{val}$ on $d$, shown in the inset of Fig.~\ref{fig:figS2}. This is because two-electron states have contributions from multiple orbitally-excited single-electron states that have different valley splittings.

\bibliography{text}

\end{document}